\documentclass[12pt,reqno]{amsart}
\usepackage{amssymb}

\newtheorem{lemma}{Lemma}

\newtheorem{comm}[lemma]{Comment}
\newtheorem{remm}[lemma]{Remark}
\newtheorem{exam}[lemma]{Example}

\newcommand{\lam}{\lambda}
\newcommand{\ome}{\omega}
\newcommand{\del}{\delta}

\newcommand{\tet}{\vartheta}

\newcommand{\Sig}{\Sigma}
\newcommand{\CA}{\mathcal{A}}

\newcommand{\DD}{\mathcal{D}}

\newcommand{\lap}{\Delta}

\newcommand{\Ss}{\mathrm{S}}

\newcommand{\R}{\mathbb{R}}

\newcommand{\T}{\mathbb{T}}
\newcommand{\Lp}[1][2]{{\rm L}^{#1}}

\newcommand{\Ic}{{\rm C}^{\infty}_{\rm c}}

\newcommand{\hull}{\mathrm{\hull}}

\newcommand{\Proof}[1][\!\!]{\textbf{\underline{Proof}.}\, #1}
\newcommand{\fin}{\hspace*{\fill} $\, \blacksquare$}

\fontshape{n}

\title[On the spectrum of a matrix model...]
{on the spectrum of a matrix model for the \emph{D}=11
supermembrane compactified on a torus with non-trivial winding.}
\author[L. Boulton, M. P. Garc\'ia del Moral, I. Mart\'\i n, A. Restuccia]{}

\date{20$^{th}$ September 2001}


\begin{document}
\




\maketitle

\centerline{\scshape L. Boulton$^1$, M. P. Garc\'\i a del
Moral$^2$, I. Mart\'\i n$^2$, A. Restuccia$^2$}

\medskip

{\footnotesize \centerline{ $^{1}$Departamento de Matem\'aticas,
Universidad Sim\'on Bol\'\i var} \centerline{Caracas, Venezuela}
 \centerline{ $^{2}$Departmento de F\'\i sica,
Universidad Sim\'on Bol\'\i var} \centerline{Caracas, Venezuela}
\centerline{\textbf{email}: lboulton@ma.usb.ve,
mgarcia@fis.usb.ve, isbeliam@usb.ve, arestu@usb.ve}}

\medskip

\begin{abstract}
The spectrum of the Hamiltonian  of the double compactified $D=11$
supermembrane with non-trivial central charge or equivalently the
non-commutative symplectic super Maxwell theory is analyzed. In
distinction to what occurs for the $D=11$ supermembrane in
Minkowski target space where the bosonic potential presents
string-like spikes which render the spectrum of the supersymmetric
model continuous, we prove that the potential of the bosonic
compactified membrane with non-trivial central charge is strictly
positive definite and becomes infinity in all directions when the
norm of the configuration space goes to infinity. This  ensures
that the resolvent of the  bosonic Hamiltonian is compact. We find
an upper bound for the asymptotic distribution of the eigenvalues.
\end{abstract}

\section{Introduction}
In \cite{dln}, the spectrum of the $\mathrm{SU}(N)$ regularized
supermembrane on $D=11$ Minkowski target space was shown to be the
whole interval $[0,\infty)$. There are two properties of the model
which render the spectrum continuous: the existence of local
string-like spikes and supersymmetry. The string-like spikes (cf.
\cite{nic2}) are local configurations with zero Hamiltonian
density of the bosonic sector. Its existence imply that the
minimal configurations of the Hamiltonian are degenerate along
some directions. The potential has the shape of valleys extending
to infinite in those directions. Consequently the bosonic theory
is classically unstable, however its associated quantum mechanical
system has discrete spectrum. There are several ways to explain
this behaviour (cf. \cite{sim}, \cite{lus} and \cite{dln}). In the
directions transverse to the valleys the system may be described
in terms of harmonic oscillators. Its zero point energy generates
a quantum effective potential which goes to infinity in all
directions when the norm in the configuration space goes to
infinite, as a consequence the quantum theory becomes stable. This
argument may be extended in a rigorous way to show that under very
mild assumptions on the potential the spectrum of the
corresponding Schr\" odinger operator becomes discrete. The
situation changes drastically when the supersymmetry is introduced
into the model. This is so, because the fermionic sector cancels
the bosonic contribution to the zero point energy. In \cite{dln}
it was proven that the spectrum of the $D=11$ supermembrane in a
Minkowski target space is continuous. We remark that the
construction in \cite{dln} is valid in the interior of an open
cone in the configuration space. A well posed spectral problem may
be defined by imposing boundary conditions to the wave functions
on the cone surface, in their case Dirichlet boundary conditions.
Its extension to the whole configuration space requires at least
the proof of existence of a well posed spectral problem in the
whole space. This problem has not been addressed in the
literature.

In this paper we will consider as in \cite{dln} wavefunctions with
support in the interior of an open cone in configuration space.
The analysis of the spectrum for the compactified supermembrane is
not yet conclusive. The main problem has been that the
$\mathrm{SU}(N)$ regularization of the compactified supermembrane
seems not to be possible. The close but not exact modes present in
the compactified case apparently do not fit into an
$\mathrm{SU}(N)$ formulation of the theory \cite{dpp}. It was,
however, pointed out there that the spectrum of the compactified
supermembrane should also be continuous, because of the presence
of string-like spikes in the configuration space of the
compactified membrane. In \cite{rus} by considering particular
configurations of the compactified supermembrane it was argued
that a non-trivial central charge should render a discrete
spectrum for the compactified supermembrane.

In \cite{MOR1} a formulation of the double compactified $D=11$
supermembrane in terms of a non-commutative geometry was obtained.
In this formulation the Hamiltonian may be expressed in terms of a
non-commutative super Maxwell theory plus the integral of the
curvature of the non-commutative connection over the world volume.
In \cite{GR} it was explicitly shown that the presence of this
curvature term introduces string-like spikes into the
configuration space of the compactified supermembrane in agreement
with \cite{dpp}. If however we restrict the theory to have a fixed
central charge, which describes a sector of the full compactified
theory then the integral of the curvature becomes zero and the
theory reduces to a non-commutative super Maxwell theory coupled
to seven scalar fields which represent the transverse directions
to the supermembrane. It was also shown in the later paper that in
this sector there are no string-like spikes. The explicit
computation was possible because in that sector of the
compactified supermembrane the closed non-exact model can be
properly handled and an $\mathrm{SU}(N)$ regularization of the
Hamiltonian may be obtained.

In section \ref{s3} we go one step forward and characterize
completely the quantum problem. We show that the potential of the
noncommutative Maxwell theory coupled to the scalar field is
strictly positive definite and becomes infinite in any direction
when the norm in the configuration space goes to infinity.
Consequently (cf. \cite{res4}) the corresponding quantum
Hamiltonian has compact resolvent so that its spectrum consists of
a discrete set of eigenvalues of finite multiplicity. The speed at
which the potential escapes to infinity is controlled by the
quadratic and quartic powers of the norm in the configuration
space, we determine upper bounds for the asymptotic distribution
of these eigenvalues by comparing with those of the harmonic
oscillator. Finally we conjecture (based on the form of the
bosonic potential) that the noncommutative supermaxwell theory
should have a discrete spectrum. This problem is currently under
investigation by some of the authors.

\section{The Hamiltonian of the double compactified supermembrane}
\label{s2} The Hamiltonian of the $D=11$ supermembrane, with
target space $M_q\times \Ss^1 \times \Ss^1$, describing
non-trivial wrapping over the compactified directions was obtained
in \cite{MOR1}, \cite{MOR2}. It was formulated in terms of a
symplectic noncommutative geometry. The resulting Hamiltonian is
exactly equivalent to the one of the $D=11$ supermembrane dual
over $M_q\times \Ss^1 \times \Ss^1$. A symplectic structure is
introduced in the formulation by considering the connection 1-form
which minimizes the Hamiltonian. The winding number $n$
characterizes a $U(1)$ bundle over the world-volume where the
minimal connection is defined. It corresponds to a monopole type
connection. The explicit expression of the bosonic Hamiltonian is
\begin{equation}\label{e5}
\begin{aligned}
 H=\int_{\Sigma}&(1/2\sqrt{W})[ (P_{m})^{2}+(\Pi_{r})^{2}+
 (1/2)W\{X^{m},X^{n}\}^{2}
 +W(\mathcal{D}_r X^{m})^{2}+\\ & +(1/2)W(\mathcal{F}_{rs})^{2}]+
 \int_{\Sigma}[(1/8)\sqrt{W}n^{2}
 -\Lambda(\mathcal{D}_{r}\Pi_{r}+\{X^{m},P_{m}\})]+\\&
 -(1/4)\int_{\Sigma}\sqrt{W}n^{*}\mathcal{F}, \qquad \qquad
 n\not=0
\end{aligned}
\end{equation}
together with its supersymmetric extension
\begin{equation}\label{e6}
\int_{\Sigma} \sqrt{W} [- \overline{\theta}\Gamma_{-} \Gamma_{r}
\mathcal{D}_{r}\theta +
 \overline{\theta}\Gamma_{-} \Gamma_{m}\{X^{m},\theta\} +
 \Lambda \{ \overline{\theta}\Gamma_{-},\theta\}]
\end{equation}
where $m=1,\ldots,7$ are indices denoting the scalar fields once
the supermembrane is formulated in the light cone gauge. They
describe the transverse directions to the world volume. The
indices $r,s=1,2$ are the ones related to the two compactified
directions of the tangent space.

We denote by $\Sig$ the spatial part of the world volume, which is
assumed to be a closed Riemann surface of genus $g$. By $P_m$ and
$\Pi_r$ we denote the conjugate momenta to $X^m$ and the
connection 1-form $\CA_r$ respectively.

By $\tet$ we denote the Majorana spinors of the $D=11$ formulation
which may be decomposed in terms of a complex 8-component spinor
of $\mathrm{SO}(7)\times \mathrm{U}(1)$.

The covariant derivative is defined by
\begin{equation*}\label{e9}
 \DD_r=\mathrm{D}_r+\{\CA_r,\,\}.
\end{equation*}
and the field strength
\begin{equation*}\label{e7}
\mathcal{F}_{rs} =\mathrm{D}_{r}\mathcal{A}_{s}-
D_{s}\mathcal{A}_{r}+ \{\mathcal{A}_{r},\mathcal{A}_{s}\}
\end{equation*}

The bracket $\{\ ,\,\}$ is defined by
\begin{equation*} \label{e8}
\{\ast,\diamond\}=
\frac{2\epsilon^{sr}}{n}(D_{r}\ast)(D_{s}\diamond )
\end{equation*}

where $n$, the winding number, denotes the integer which
characterizes the non trivial $U(1)$ principle bundle over $\Sig$.

In the above, $\mathrm{D}_{r}$ is a tangent space derivative
defined by
\begin{equation*} \label{e10}
 \mathrm{D}_{r}=
 \frac{\widehat{\Pi}^{a}_{r}}{\sqrt{W}}\partial_{a}
\end{equation*}
where $\partial_{a}$ denotes derivatives with respect to the local
coordinates on $\Sig$. By $\widehat{\Pi}^a_{r}$ we denote a
zwei-vein defined from the minimum of the Hamiltonian (\ref{e5}).
It satisfies
\begin{equation*}\label{e11}
\begin{gathered}
 \widehat{\Pi}^a_{r} =\epsilon^{ab}\partial_b \widehat{\Pi}_{r} \\
 \left\{ \widehat{\Pi}_{r},\widehat{\Pi}_{s} \right\} =
 (1/2)n \epsilon_{rs}.
\end{gathered}
\end{equation*}
In \cite{GR} an $\mathrm{SU}(N)$ regularization of (\ref{e5}),
(\ref{e6}), for the case in which the symplectic connection
$\CA_r$ has no transitions over $\Sig$, was obtained. In that case
the latter term of (\ref{e5}), which is the integral of a total
derivative of a single-valued object over $\Sig$, vanishes.

The resulting $\mathrm{SU}(N)$ model is
\begin{equation}\label{e12}
\begin{aligned}
H= & \mathrm{Tr}\left(\frac{1}{2N^{3}}(P^{0}_mT_{0}P^{0}_{m}T_{0}+
\Pi_r^0T_{0}\Pi^{-0}_{r}T_{0}+(P_{m})^2+ (\Pi_{r})^{2})+
\right.\\& +\frac{n^2}{16\pi^2N^3}[X^{m},X^{n}]^2+
\frac{n^2}{8\pi^2N^3}\left(\frac{i}{N}[T_{V_{r}},X^{m}]T_{-V_{r}}-
[\mathcal{A}_r,X^{m}]\right)^2+\\&
+\frac{n^2}{16\pi^2N^3}\left([\mathcal{A}_r,\mathcal{A}_s]+
\frac{i}{N}([T_{V_s},\mathcal{A}_r]T_{-V_s}-[T_{V_r},\mathcal{A}_s]
T_{-V_r})\right)^2 + \frac{1}{8}n^2+\\& +\frac{n}{4\pi N^3}
  \Lambda\left([ X^{m},P_{m}]- \frac{i}{N}[T_{V_r},\Pi_{r}]T_{-V_r}
  +[ \mathcal{A}_{r},\Pi_{r}]\right)+\\
   &+ \frac{in}{4\pi N^3}(\overline{\Psi}\gamma_{-}\gamma_{m}
   \lbrack{X^{m},\Psi}\rbrack
   -\overline{\Psi}\gamma_{-}\gamma_{r}\lbrack{\mathcal{A}_{r},\Psi}
   \rbrack +
  \Lambda \lbrack{\overline{\Psi}\gamma_{-},\Psi}\rbrack + \\
  & - \left.\frac{i}{N} \overline{\Psi}\gamma_{-}\gamma_{r}
  [T_{V_{r}},\Psi] T_{-V_r})\right)
\end{aligned}
\end{equation}
subject to
\begin{equation}\label{e13}
\begin{aligned}
\mathcal{A}_{1}= &\mathcal{A}^{(a_1,0)}_{1}T_{(a_1,0)}, \\
\mathcal{A}_{2}= &\mathcal{A}^{(a_1,a_{2})}_{2} T_{(a_1,a_2)}
\quad \textrm{with}\quad a_2\neq0.
\end{aligned}
\end{equation}
We use the following notation
\begin{equation*} \label{e14}
\begin{gathered}
 X^{m}=  X^{mA}T_{A}\quad\qquad P_{m}=P{A}_mT_{A}\\
 \mathcal{A}_r= \mathcal{A}^{A_{r}} T_{A} \quad\qquad
 \Pi_{r}=\Pi_r^{A}T_{A}
\end{gathered}
\end{equation*}
where $T_A$ are the generators of the $\mathrm{SU}(N)$ algebra:
\begin{equation*} \label{e15}
 [T_{A},T_{B}]= f^{C}_{AB}T_{C}.
\end{equation*}
It was shown in \cite{GR} that this regularized Hamiltonian has an
associated mass operator with no string-like spikes. That is, the
local conditions on the bosonic sector
\begin{gather}
 \begin{aligned}
 \mathrm{Tr} \left[\frac{n^2}{16\pi^2N^3}\right.& [X^{m},X^{n}]^2+
 \frac{n^2}{8\pi^2N^3}\left(\frac{i}{N}[T_{V_{r}},X^{m}]T_{-V_{r}}-
 [\mathcal{A}_r,X^{m}]\right)^2+  \\&
 \left.+\frac{n^2}{16\pi^2N^3}\left([\mathcal{A}_r,\mathcal{A}_s]+
 \frac{i}{N}([T_{V_s},\mathcal{A}_r]T_{-V_s}-[T_{V_r},\mathcal{A}_s]
 T_{-V_r})\right)^2\right]=0,
 \end{aligned}  \label{e16} \\
 [ X^{m},P_{m}]- \frac{i}{N}[T_{V_r},\Pi_{r}]T_{-V_r}
  +[ \mathcal{A}_{r},\Pi_{r}]=0, \label{e17}\\
 \CA_1=\CA_1^{(a_1,0)}T_{(a_1,0)} \label{e18}\\
 \CA_2=\CA_2^{(a_1,a_2)}T_{(a_1,a_2)}, \qquad \mathrm{with} \quad
 a_2 \not=0, \label{e19}
\end{gather}
imply
\begin{equation}\label{e20}
  X^{Bm}=0,\qquad \CA^B_r=0, \qquad P^B_m=0, \qquad \Pi^B_r=0.
\end{equation}

The constraint (\ref{e17}) determines $\Pi_1^{(a,b)},b\not=0$ and
$\Pi_2^{(a,0)}$ which together with (\ref{e18}) and (\ref{e19})
allow a canonical reduction $H_R$ for the Hamiltonian (\ref{e12}).
The same canonical reduction may be performed in (\ref{e6}) when
the geometrical objects are expressed in a complete orthonormal
basis of the space $\Lp(\Sig)$. After this reduction, the term
$|\Pi^{(a,b)}_1|^2,b\not =0$ and $|\Pi_2^{(a,0)}|^2$ become
non-trivial, however since they are positive, we can bound the
mass operator
\[
 \mu_R=H_R-\mathrm{Tr} \left(\frac{1}{2N^3} P_m^{0}T_0P^0_mT_0+
  \Pi^0_{r} T_0\Pi_r^0 T_0 \right)
\]
by an operator $\mu$ without such terms. If the resulting $\mu$ is
bounded from below and has a compact resolvent, the same
properties are valid for $\mu_R$. We will show the former in the
forthcoming section. Notice that $\mu$ is of Schr\"odinger type
with potential the left hand side of (\ref{e16}). In what follows
it will be understood without further mention that the centre of
mass terms have been removed. This restrictions \eqref{e13}
associated to gauge fixing conditions of \eqref{e5}, \eqref{e6}
are equivalent to the ones used in \cite{dln}. In their analysis
they use the gauge freedom to diagonalize one of the $X$ maps, the
residual gauge freedom when $X$ is regular consists of an
arbitrary element of the Cartan subgroup.

The imposition of \eqref{e12} together with the elimination of the
associated conjugate momenta is valid only in the interior of an
open cone $K$. Our model thus considers as in \cite{dln},
wavefunctions with support on the interior of $K$. For technical
reasons, in order to show that the spectrum of $\mu$ is discrete,
in the following section we also consider $\mu$ as an operator
acting on the whole configuration space. We shall see that
discreteness of the spectrum for the latter implies the same
property for the restriction to any hyper-cone.

\section{Discretness of the spectrum of $\mu$} \label{s3}
We introduce the following notation. The Hamiltonian
\[
  \mu=-\lap_X -\lap_\CA + V(X,\CA)
\]
acts on $\Lp((X,\CA)\in \R^M)$ for suitably large $M$. The
potential $V$ is given by
\[
 V(X,\CA) = V_1(X) + V_2(\CA) + V_3(X,\CA)
\]
where
\[
   V_1(X)= 4\sum_{D,m,n} \left|N \sin \left(\frac{B\times C}{N} \pi\right)
   X^{Bm}X^{Cn} \del ^D_{B+C}\right|^2,
\]

\begin{align*}
   &V_2(\CA)=  4\sum_{r,s,D} \left|\ome^{(D\times V_r)/2}
   N \sin \left(\frac{V_r\times D}{N}
   \pi\right)\CA_s^{D} + \right. \\ &\left. -\ome^{(D\times V_s)/2}N
   \sin \left(\frac{V_s\times D}{N} \pi\right)
   \CA_r^D + i
   \sum_{B,C} N \sin \left(\frac{B\times C}{N} \pi\right)
   \CA_r^B \CA_s^C \del_{B+C}^D \right|^2,
\end{align*}

\begin{align*}
   V_3(X,\CA)=  2 \sum_{D,r,m}&\left|\ome^{(D\times V_r)/2}
   N \sin \left(\frac{V_r\times D}{N}
   \pi\right) X^{Dm} \right.\\ & \left.+ i \sum_{B,C}
   N \sin \left(\frac{B\times C}{N} \pi\right)\CA^B_rX^{Cm} \del
   ^D_{B+C}\right|^2,
   \end{align*}

\[
   \ome=\exp\left(\frac{2\pi i}{N}\right).
\]

We define rigorously $\mu$ as the self-adjoint non-negative
Friedrichs extension of $(\mu,\Ic(\R^M))$. Our aim is to show that
$\mu$ has compact resolvent. The proof depends upon the fact that
$V$ is a basin shape potential.

As we mentioned in section \ref{s2}, the operator $\mu$ that
bounds our model \eqref{e5} should not be defined on the whole
space $R^M$ but only on the open hyper-cone $K\subset \R^M$. Every
sequence of approximate eigenfunctions squared integrable on $K$
is also a sequence of approximate eigenfunctions squared
integrable on the whole space. Therefore, if the spectrum of $\mu$
as an operator acting on $\R^M$ is discrete, it also has spectrum
discrete as an operator acting on $K$. We use this without further
mention. Notice that the same argument applies to any open
hyper-cone on the configuration space.

\medskip

\begin{lemma} \label{t1}
The potential $V(X,\CA)=0$, if and only if $X^{Bm}=0$ and
$\CA^B_r=0$ for all indices $B,m,r$.
\end{lemma}
\Proof The condition $V_2(\CA)=0$, yields
\begin{equation*}\label{e21}
 k^{-(V_{r}\times
 A)/2}\widetilde{\lambda}_{rA}\mathcal{A}^{A}_{s}-
 k^{-(V_{s}\times
 A)/2}\widetilde{\lambda}_{sA}\mathcal{A}^{A}_{r}+f^{A}_{BC}
 \mathcal{A}^{B}_{r}\mathcal{A}^{C}_{s}=0.
\end{equation*}
Using (\ref{e18}) and (\ref{e19}) we obtain
\begin{equation*}\label{e22}
\begin{gathered}
 (1/2) k^{-(V_{1}\times A)/2}N\sin \left(\frac{V_{1}\times
 A}{N}\pi\right)\mathcal{A}^{A}_{2}- k^{-(V_{2}\times A)/2}N \sin
 \left(\frac{V_{2}\times A}{N}\pi\right)\mathcal{A}^{A}_{1}+\\
 + i N\sin \left(\frac{b_1V_{1}\times
 A}{N}\pi\right)\mathcal{A}^{b_1,0}_{1}\mathcal{A}^{A-b_1V_{1}}_{2}=0
\end{gathered}
\end{equation*}
where the $b_i$ are integers. In particular for $\CA=lV_1$, $l$
integer, we get $\CA^{lV_1} \equiv \CA ^{l,0}_1 =0$ hence
$\CA^A_1=0$. We then obtain from (\ref{e22}) and (\ref{e19}),
$\CA_2^A=0$. The condition
\[
 V_3(X,\CA)=0
\]
then reduces to
\begin{equation*} \label{e23}
 \begin{gathered}
 k^{(A\times V_1)/2} \sin \left(\frac{V_1\times A}{N}\pi \right)
 X^{mA} =0, \\
 k^{(A\times V_2)/2} \sin \left(\frac{V_2\times A}{N}\pi \right)
 X^{mA} =0,
 \end{gathered}
\end{equation*} which yields $X^{mA}=0$. \fin

\medskip

The following proposition is the main result of this section.

\begin{lemma} \label{t2}
The potential $V(X,\CA)\to \infty$ as $(X,\CA) \to \infty$.
\end{lemma}
\Proof Let $\T$ be the unit ball of $\R^M$. We write $X^{Bm}$ and
$A^B_r$ in polar coordinates as
\[
 X^{Bm}=R\phi^{Bm}, \, \CA^B_r=R \psi^B_r \qquad \mathrm{or}
 \qquad X=R\phi, \, \CA=R \psi
\]
where $R\geq 0$, $\phi=(\phi^{Bm})$, $\psi=(\psi^B_r)$ and
$(\phi,\psi)\in \T$. In order to show the desired limit, we shall
show
\begin{equation} \label{e1}
 \inf_{(\phi,\psi) \in \T}V(R\phi,R\psi) \to \infty
 \qquad \mathrm{as} \qquad R\to \infty.
\end{equation}

Elementary computations yield
\[
  V(R\phi,R\psi)=R^4k_1(\phi,\psi)+R^3k_2(\phi,\psi) +
  R^2k_3(\phi,\psi),
\]
where $k_1(\phi,\psi) \geq 0$, $k_3(\phi,\psi) \geq 0$,
$k_2(\phi,\psi)\in \R$ can be negative and if $k_1(\phi,\psi)=0$,
then $k_2(\phi,\psi)=0$. The $k_j$ are continuous in
$(\phi,\psi)\in \T$ and by virtue of lemma \ref{t1},
\[
   K:= \inf_{(\phi,\psi) \in \T}
   [k_1(\phi,\psi)+k_2(\phi,\psi)+k_3(\phi,\psi)]=
   \inf_{(\phi,\psi) \in \T} V(\phi,\psi) >0.
\]
When clear from the context, below we will write $k_j\equiv
k_j(\phi,\psi)$.

 Let
\begin{gather*}
 \T_1= \{(\phi,\psi)\in \T\,:\, k_1(\phi,\psi)=0 \} \\
 \T_2= \{(\phi,\psi)\in \T\,:\, k_1(\phi,\psi)>0 \}.
\end{gather*}
Then $\T=\T_1\cup \T_2$ and since $k_1$ is continuous
$\overline{\T_2}=\T$. Put
\[
   V(R\phi,R\psi)=R^2(R^2k_1(\phi,\psi)+Rk_2(\phi,\psi)+
   k_3(\phi,\psi)) =R^2P_{\phi,\psi}(R).
\]
Then $P_{\phi,\psi}(R)$ is a family of paraboles parameterized by
$(\phi,\psi)\in \T$ such that
\begin{gather*}
 P_{\phi,\psi}(0)=k_3(\phi,\psi), \\
 P''_{\phi,\psi}(R)=k_1(\phi,\psi) \geq 0 \qquad \mathrm{and}\\
 P_{\phi,\psi}(1)=k_1(\phi,\psi)+k_2(\phi,\psi)+k_3(\phi,\psi)
 \geq K > 0.
\end{gather*}

Now
\begin{equation} \label{e2}
 \inf_{(\phi,\psi)\in\T} V(R\phi,R\psi)=
 R^2 \inf P_{\phi,\psi}(R) \geq R^2
 \inf_{(\phi,\psi)\in\T} M(\phi,\psi),
\end{equation}
where
\[
 M(\phi,\psi):= \min_{R\geq 1} P_{\phi,\psi}(R).
\]
If $M:\T \longrightarrow \R$ was a strictly positive continuous
function, since $\T$ is compact necessarily
\begin{equation} \label{e24}
 \inf_{(\phi,\psi)\in\T} M(\phi,\psi) \geq \tilde{M}>0,
\end{equation}
thus the right hand side of (\ref{e2}) escapes to $\infty$ as
$R\to \infty$ and so (\ref{e1}) is proven.

In order to complete the proof, we show first that $M$ is a
strictly positive function. If $(\phi,\psi)\in\T_1$, then $k_1$
and hence $k_2$ vanish so $M(\phi,\psi)=k_3(\phi,\psi)>0$. On the
other hand, if $M(\phi,\psi)=0$ for some $(\phi,\psi)\in\T_2$, we
would have $P_{\phi,\psi}(R_0)=0$ for some $R_0>0$ so that
\[
  V(R_0\phi,R_0\psi)=R_0^2P_{\phi,\psi}(R_0)=0
\]
hence we contradict lemma \ref{t1}. Therefore necessarily
$M(\phi,\psi)>0$ for all $(\phi,\psi)\in \T_2$.

Finally we show that $M$ is continuous. For this we consider
separately the regions $\T_1$ and $\T_2$. If $(\phi,\psi)\in
\T_2$, the minimum of the polynomial $P_{\phi,\psi}$ is attained
at
\[
   R_0\equiv
   R_0(\phi,\psi)=-\frac{k_2}{2k_1}.
\]
At this point
\[
  P_{\phi,\psi}(R_0)=k_3-\frac{k_2^2}{4k_1^2}.
\]
If $k_2\geq -2k_1$, then $R_0<1$ so
\[
   M(\phi,\psi)=P_{\phi,\psi}(1)=
   k_1+k_2+k_3.
\]
If $k_2<-2k_1$, then
\[
 M(\phi,\psi)=k_3-\frac{k_2^2}{4k_1^2}.
\]
In both cases the continuity of $M$ at $(\phi,\psi)\in \T_2$
follows from the continuity of the $k_1,k_2,k_3$ and the fact that
$k_1(\phi,\psi)\not=0$.

Since $\T_1 \subset \overline{\T_2}$, in order to show the
continuity of $M$ in $\T_1$, it is enough to prove that for any
sequence $(\phi_n,\psi_n)\in \T_2$, such that
\[
  (\phi_n,\psi_n)\to (\phi,\psi)\in \T_1 \qquad \mathrm{as}
  \qquad n\to \infty,
\]
we also have
\begin{equation} \label{e3}
  M(\phi_n,\psi_n) \to M(\phi,\psi).
\end{equation}
For this let the family of lines
\[
  L_n(R)=[k_1(\phi_n,\psi_n)+k_2(\phi_n,\psi_n)]R+
    k_3(\phi_n,\psi_n) \qquad R>0.
\]
Since
\[
P''_{(\phi_n,\psi_n)}(R)>0, \, L_n(0)=P_{\phi_n,\psi_n}(0)
\]
and
\[
 L_n(1)=P_{\phi_n,\psi_n}(1),
\]
one has
\[
 L_n(R)\leq P_{\phi_n,\psi_n}(R) \qquad \mathrm{for\ all}
 \qquad R\geq 1.
\]
Then for all $n$ large
\begin{equation} \label{e4}
 L_n\left(\frac{1}{k_1^2(\phi_n,\psi_n)}\right) \leq
 P_{\phi_n,\psi_n}\left(\frac{1}{k_1^2(\phi_n,\psi_n)}\right).
\end{equation}
Since $k_1(\phi_n,\psi_n)\to 0$ as $n\to\infty$,
\[
   k_1(\phi_n,\psi_n)+k_2(\phi_n,\psi_n)+k_3(\phi_n,\psi_n) \geq K
\]
and the $k_2(\phi_n,\psi_n)$ are bounded, both the left hand side
and the right hand side of (\ref{e4}) tend to $k_3(\phi,\psi)$ as
$n \to \infty$. Hence the family of straight lines $L_n(R)$
approaches to the horizontal line $y\equiv k_3(\phi,\psi)$ as
$n\to \infty$. An elementary geometric argument (notice
$-k_2/(2k_1)<1/k_1^2$ for all large $n$) confirms that this is
enough to guarantee (\ref{e3}) and so the proof is complete. \fin

\medskip

As a consequence of lemma \ref{t2} and from a standard result in
the spectral theory of Schr\"odinger operators (cf. \cite{res4}),
the resolvent of $\mu$ is compact so the spectrum consists purely
of isolated eigenvalues of finite multiplicity. By the positivity
of $V$, we also know that all such eigenvalues are non-negative.

Let us conclude this section by describing how the estimates in
the proof of lemma \ref{t2} provide information about the
distribution of the eigenvalues of the original model $H$. For
simplicity we introduce the following standard notation. If the
linear operator $T$ is bounded below and has discrete spectrum, we
define $N_T(\lam)$ (the counting function of $T$) as the number of
eigenvalues of $T$ counting multiplicity which are less or equal
to $\lam>0$.

We find upper bounds for $N_H(\lam)$ as follows. Each eigenvalue
of $\mu$ as an operator acting on an hyper-cone will also be an
eigenvalue of $\mu$ as an operator acting on the whole space
$\R^M$, thus the counting function of the former will always be
below the counting function of the latter. Since $H$ only differs
from $\mu$ (acting on $K$) by a positive term, the above yields
\[
 N_H(\lam)\leq N_\mu(\lam)
\]
where now the $\mu$ at the right hand side acts on $\R^M$.

Explicit estimates on the eigenvalues of $\mu$ come from bounds on
the potential. By virtue of \eqref{e2} and \eqref{e24}, there
exist constants $\tilde{M}>0$ and $b>0$ such that
\[
 V(R\phi,R\psi) \geq \tilde{M} R^2-b \qquad \qquad \phi,\psi\in
\T,\, R>0.
\]
According to the min-max principle (cf. \cite{res4}), the
eigenvalues of $\mu$ are bounded from below by those of the
harmonic oscillator
\[
 \tilde{\mu}=-\lap_X -\lap_\CA + \tilde{M}(|X|^2+|\CA|^2)-b.
\]
This yields $N_{\mu}(\lam) \leq N_{\tilde{\mu}}(\lam)$. Now, the
spectrum of $\tilde{\mu}$ can be computed explicitly, indeed
$N_{\tilde{\mu}}(\lam)$ is of the order of $\lam$ as $\lam \to
\infty$. By collecting all these estimates, we gather that
\[N_H(\lam) = O(\lam)\] as $\lam\to \infty$.

We should mention that the leading term in $V(X,\CA)$ is not
quadratic but quartic. This suggests that the estimate found above
is not very sharp. At the present moment an investigation on
better bounds for the eigenvalues of $H$ is being carried out by
some of the authors.

\bigskip

{\samepage {\scshape Acknowledgments.} The authors wish to thank
the Department of Mathematics at Kings College London for their
kind hospitality while part of this work was done.}

\end{document}